# First broadband optical fibre with an attenuation lower than 0.1 decibel per kilometre

Marco Petrovich[1]*, Eric Numkam Fokoua[1]*, Yong Chen[1], Hesham Sakr[1], Abubakar Isa Adamu[1], Rosdi Hassan[1], Dong Wu[1], Ron Fatobene Ando[1], Athanasios Papadimopoulos[1], Seyed Reza Sandoghchi[1], Gregory Jasion[2], Francesco Poletti[1,2,†]

* These authors contributed equally to this work

 1. Microsoft Azure Fiber, Unit 7, The Quadrangle, Romsey, S051 9DL, UK.

2. Optoelectronics Research Centre, University of Southampton, SO17 1BJ, UK,

† *email: fp@soton.ac.uk*

## Abstract

Throughout history, the development of novel technologies for long-distance communications has had profound influences on societal progress. Landmark scientific discoveries have enabled the transition from short message transmissions via single-wire electrical telegraphs to voice communications through coaxial cables, and ultimately to the optical fibres that powered the internet revolution. Central to these advancements was the invention of novel waveguides to transport electromagnetic waves with lower attenuation over broader spectral ranges. In the past four decades, despite extensive research, the spectral bandwidth and attenuation of silica-based telecommunication fibres have remained relatively unchanged. In this work, we report an optical waveguide with an unprecedented bandwidth and attenuation. Its measured loss reaches 0.091 dB/km at 1550 nm and remains below 0.2 dB/km over 66 THz, substantially better than the 0.14 dB/km and 26 THz achievable with existing technology. Our innovative, microstructured optical fibre replaces the traditional glass core with air, employing a meticulously engineered tubular glass structure to guide light. This approach not only reduces attenuation and other signal degradation phenomena, but it also increases transmission speeds by 50%. Furthermore, it theoretically enables further loss reductions and operation at wavelengths where broader bandwidth amplifiers exist, potentially heralding a new era in long-distance communications.

# 1. Introduction

The quest for long-distance communication has driven human creativity for centuries, from the use of fire beacons at night in the Old and Middle Ages, to the mechanical optical telegraphs of the Napoleonic era, up to the groundbreaking electric telegraphs of the 1850s. The transmission of the first Morse-coded message across the Atlantic via a sub-sea telegraph cable in 1858 was a monumental achievement that shrank geographical divides and revolutionized communication[1]. The realisation in the early 20th century that modulated radio waves could be reflected by the ionosphere further enhanced communication capabilities, thus enabling long distance communications even in the absence of a direct connection and of a line of sight[2]. However, the inherent noisiness, unreliability and limited bandwidth of radio wave communication prompted the development of higher-quality cables that could transmit multiple voice calls simultaneously. Heaviside's coaxial cable[3], with suitably developed conductive and insulating materials, became the technology that underpinned long distance transmissions for decades. The transition from coaxial cables to optical fibres marked another significant milestone in communication technology. The pioneering work of Kao and Hockham in the 1960s identified the potential of using purified glass for transmitting modulated optical signals (hence information) to kilometre-scale distances[4], leading to the development of low-loss optical fibres by Corning in the 1970s[5]. This innovation ushered in the era of digital optical communications, which for the last half a century has formed the backbone of global telecommunication networks and enabled the internet revolution[6]. Is a further step ahead possible?

All these breakthroughs were driven by the primary objective to transmit more information, as either more simultaneous messages and voice calls in the analogue electrical era, or more bits per second in the digital age. A second, non-negligible goal has always been the reduction of the attenuation (or "loss") of the transmission medium, to increase the distance a signal could reach before needing regeneration or amplification. Shannon's mathematical theory of information linked the two goals: lower attenuation required less amplification; the resulting improvement in signal to noise ratio enabled the system to increase its maximum throughput of information[7].

Upshifting the frequency of the modulated signal carrier from tens of MHz used in the long-distance electrical coaxial cables to hundreds of THz used in optical communications enabled an increase in information throughput of more than a million times. Simultaneously, optical fibres also presented an ultralow level of attenuation of around 0.15 dB/km, which remained approximately constant over a bandwidth of ~10 THz where optical amplification from Erbium Doped Fibre Amplifiers (EDFAs was available. This was a significant improvement over coaxial cables, where attenuation was frequency dependent (as $\sqrt{f}$ ) and reached much higher values than optical fibres at the top frequencies (e.g. ~4.5 dB/km at 30 MHz in the transatlantic TAT-6 cable[8]).

Despite unrelented progress in the field of optical communications since 1970, the minimum attenuation of silica glass fibres has remained approximately unchanged for more than four decades: from 0.154 dB/km in 1985[9] to 0.1396 dB/km in 2024[10]. The seemingly insurmountable attenuation limit of ~0.14 dB/km for information-carrying waveguides has so far hindered further breakthroughs in communication systems. It has also forced technology to converge to this relatively narrow frequency range of only 5% of the carrier frequency (10 THz at around 192 THz).

Having failed in many decades to identify and synthetise a more transparent glass than silica, a potential route to further lower the propagation loss of a long-distance communication waveguide is to *avoid* the scattering and absorptions introduced by the glass and which cause loss of signal power in telecoms fibres This can be achieved by transmitting electro-magnetic radiation in a *hollow region* rather than through a solid glass core. Theoretical foundations[11], early loss estimates[12] and first experiments[13,14] for cylindrical, metal, hollow waveguides pre-dated the development of ultra-pure glass fibres. Experimental works from Bell Labs in the mid-20th century with dielectric coated metallic hollow pipes (WT4) reached losses as low as 0.5 dB/km at frequencies of 70 GHz and impressive capacities of 476,000 voice channels[15]. The technology was however discarded in the mid-1970s for installation complexities and techno-economic reasons.

New research in the late1990s and 2000s investigated the potential for achieving ultralow loss at visible/near infrared frequencies by transmitting light through hair-thin flexible hollow core fibres (HCFs). These glass-based waveguides could transmit light in an air core, thanks to a periodic 'holey' cladding around it that created an out-of-plane photonic bandgap[16–19]. Whilst such research produced an outstanding new tool for scientific investigations, it failed to attain fibres with attenuation below 1 dB/km and with adequate modal purity for long distance communication. It is only with the advent of a second generation of HCFs, guiding light through antiresonances and inhibited coupling effects in sub-wavelength-thick, core-surrounding membranes[20], and with the introduction of nested tube designs[21,22], that the prospect of achieving sub 0.14 dB/km losses became viable[22]. Over the last 6 years, through improved designs and engineering, loss in these Nested or Double Nested Antiresonant Nodeless hollow core fibres (NANFs/DNANFs) has decreased by an order of magnitude, reaching near parity with the fundamental attenuation of silica glass telecoms fibres at 1550 nm[23], and lower values at both shorter[24–26] and longer[27] wavelengths.

In this work, we showcase the latest advancements in hollow core DNANF technology and present the first optical waveguide that surpasses conventional optical fibres in both loss and bandwidth simultaneously. With a measured loss of under 0.1 dB/km across an 18 THz bandwidth, this breakthrough result paves the way for a potential revolution in optical communications, enabling unprecedented data transmission capacities, more energy-efficient optical networks, and longer unamplified spans.

## 2. Fibre fabrication and characterisation

Optimising the loss of hollow core DNANFs requires in-depth understanding and accurate models of all its three primary components: leakage loss (LL), surface scattering loss (SSL) and microbend loss (µBL). Each mechanism has a different dependence on the geometrical features of the fibre's cross section and on the operating wavelength[28]. SSL and µBL models require statistical characterisation of the surface roughness of the microstructure's glass membranes and of the micro perturbations external to the fibre, respectively, that cannot be acquired with adequate accuracy at all spatial frequencies. To circumvent the problem, we fit the free parameters in well-established SSL and µBL models[28] so that good match between measured and simulated loss is achieved simultaneously for a statistically significant number of different DNANFs fabricated according to our processes[29]. Fig. 1 shows the excellent overall agreement between the measured loss of a DNANF, randomly chosen amongst the fifteen different fibres we used to fit our models, and our modelling prediction. As can be seen, the three loss mechanisms have notably different spectral behaviours, but, when added together, they reproduce accurately the measured fibre loss, not only in the fundamental antiresonant window (~1300 to >1700 nm) where SSL provides the largest contribution, but also in the second window (~750-840 nm) where µBL dominates.

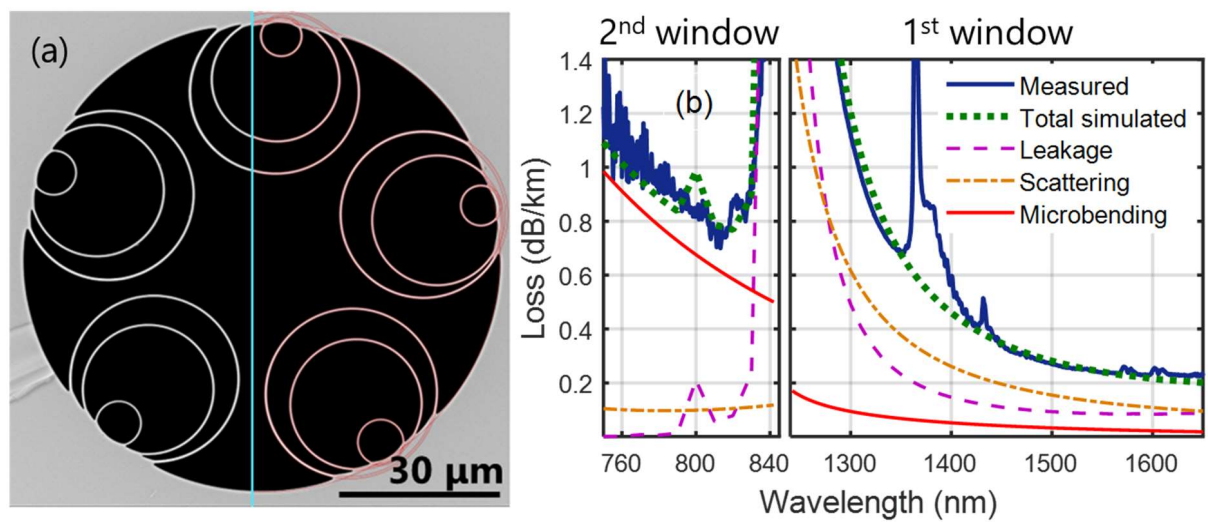

**Fig.1: Modelling validation.** (a) Scanning Electron Microscope (SEM) image of the cross-section of one of the DNANFs used to calibrate our loss model. In red the automatic edge extraction of the tubular resonators used to inform simulations; (b) measured loss of the same fibre and simulation of its LL, SSL, µBL. The total simulated loss shows good agreement with the measurement in both the first (left) and second (right) antiresonant windows, despite being dominated by different mechanisms.

We then used this calibrated loss model to optimize the fibre geometry. To produce the widest possible low-loss bandwidth, we targeted a fibre in which the (widest) first antiresonance window is centred at 1550 nm. Besides, we also aimed to achieve sufficiently high modal purity to yield an intermodal interference (IMI) of better than -60 dB/km, adequate to maintain a negligibly low penalty in long haul distance transmissions[30,31]. Modelling predictions indicate that for core sizes between 25 and 36 μm, total losses below 0.1 dB/km are possible, with the dominant contribution shifting from SSL to μBL as the core enlarges. In this work we settled on a core diameter of ~29 μm, theoretically capable of achieving a loss as low as 0.07 dB/km.

Driven by modelling guidance, we fabricated a first fibre (HCF1) with a core diameter of 28.8±0.5 μm and average diameters of the nested tubes of 31.0±1.5 μm (large tubes), 28.8±2 μm (middle) and 10.0±3 μm (small). The membrane thicknesses for all tubes were around 500 nm, tailored to centre the fundamental antiresonance window at 1550 nm, and the fibre length is 4.12 km[29]. We used three different methods to characterise its loss. As predicted by modelling, this was measured to be lower than that of glass-core fibres. However, due to the short fibre length available and its record-low level of loss, despite extreme care in performing the measurements three different techniques yielded rather different loss values. At 1550 nm, two independent cutbacks averaged a loss of 0.055 and 0.065 dB/km; through an insertion loss measurement we obtained 0.09±0.01 dB/km, while with an optical time-domain reflectometer (OTDR) we measured 0.11±0.01 dB/km[29]. This clearly indicates that significantly longer fibres are needed for accurate measurements.

After our disclosure[29], another team reported fabrication of a DNANF variant that achieved similar loss values, in the range of 0.1 to 0.13 dB/km and very high suppression of high order modes. The fibre had four rather than five sets of double nested tubes, semicircular outer tubes and a core size in the range of 29 to 31 μm[32]. Its fabricated lengths were similar to our HCF1, with five bands between 2.3 and 4.2 km. The thicker membranes (~1.1 vs ~0.5 μm of our work) make fabrication easier at the expense of a considerably narrower bandwidth. While such a fibre concept might be a good candidate for data transmission in the C or C+L telecommunication bands, it is intrinsically unable to support the ultrawide band operation that we seek in this work.

We then worked on upscaling the fabricated length of our first-window DNANF concept and produced HCF2. The fibre cross section is shown in Fig. 2(a). Its geometry closely resembles that of HCF1, with a core diameter varying between 29.1 and 29.6 μm across the full length. The diameters of the nested tubes vary azimuthally and longitudinally between 30.4 and 31.7 μm (large), 22.7 and 24.8 μm (middle) and 7.0 and 8.4 μm (small). The fibre length is 15 km, which allows for substantially improved accuracy in its loss measurement. Figure 2(b) shows OTDR measurements at 1310 and 1550 nm, indicating a uniform loss with a slope of 0.123 and 0.095 dB/km, respectively. To validate these measurements, we performed four separate cutbacks on the fibre, using two different methods. The green band in Fig. 2(c) shows the 95% confidence level of these four measurements around the mean loss curve, shown in dark blue (see Methods). We also performed an additional cutback using an Optical Spectrum Analyser (OSA) capable of measuring wavelengths in the 1650-1900 nm range and of capturing the full spectral transmission bandwidth of the fibre (light blue curve in Fig. 2(c)). As can be seen in the zoomed-in plots of Fig. 2(d), the average loss is 0.128 and 0.091 dB/km at 1310 and 1550 nm respectively. This is in good agreement with the OTDR measurements, which fall within the confidence level of the cutback measurements. The loss is below 0.1 dB/km between 1481 and 1625 nm. Remarkably, a lower loss than the current 0.14 dB/km record of solid-core fibres is also achieved in the O-band, between 1292 and 1347 nm.

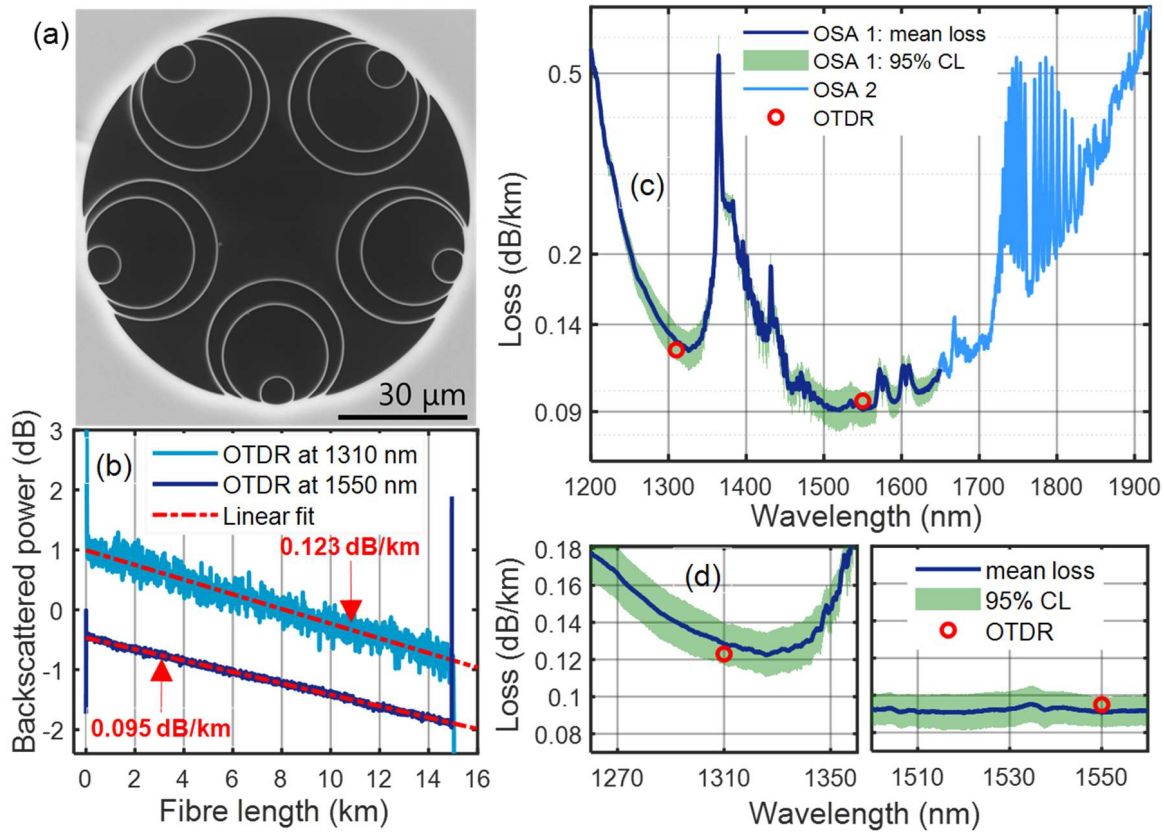

**Fig.2: Characterisation of the fabricated fibre.** (a) SEM of the fabricated HCF2; (b) bi-directional OTDR traces at 1310 and 1550 nm; (c) mean loss and 95% confidence level of four separate cutbacks (1200-1650 nm) compared to OTDR results and a fifth cutback to show the spectral loss in the 1650-1950 nm range; (d) cutback loss details in the 1270-1350 nm and 1500-1560 nm bands.

The loss curve in Fig. 2(c) shows signs of absorptions from gaseous species present in the core of the fibre. Comparison with data from the HITRAN database[33] allows one to identify absorptions from water vapour, carbon dioxide and some nitrogen oxides, likely to arise from residual atmospheric gases in the fibre's preform, as well as hydrogen chloride originating from the use of chlorinated glass tubes to make the fibre. None of these absorptions is a fundamental property of the fibre, and improved fabrication processes are expected to lead to a reduction or elimination of these absorption features. For example, use of glass with no Chlorine would straightforwardly eliminate the HCl lines and open an 18.5 THz absorption-free window between 1620 and 1800 nm, where the low loss region could be tuned (see later). $CO_2$, water vapour and $NO_x$ gases are unintentionally added to the hollow regions of the fibre during the fabrication and could be substantially reduced through process improvements. Besides, all these absorption lines have a narrow linewidth (of the order of a few picometres) and occur at known frequencies, which makes it possible to foresee the development of transmission strategies to compensate for their presence.

Besides the loss, we also measured the intermodal interference (IMI) and the polarisation properties of the fibre around 1550 nm. IMI was measured to be as low as -70dB/km, which according to previous studies would add no measurable penalty for any long-distance data transmission[30,31]. The fibre's polarisation dependent loss (PDL) and polarisation mode dispersion (PMD) coefficients are 0.013 dB/√km and 0.1 ps/√km, respectively, not too dissimilar from those of standard telecoms fibres.

Figure 3(a) shows how the mean loss of HCF2 compares with that of the pure silica core fibre (PSCF) with the lowest loss reported to date (Sato et al, 0.1406 dB/km at 1550 nm[10]). For visual comparison, we also show the loss curve of the record-low loss fibre in 2002 from Nagayama et al.[34]. In this fibre the loss was measured across a broad wavelength range (1250-1750 nm), which helps comparing the fundamental spectral bandwidth of silica versus that of HCF2. The difference between Nagayama's and Sato's fibres also shows the progress of solid core research fibres in the last 23 years. As can be seen, neglecting gas absorptions, HCF2 has a fundamental attenuation of less than 0.14 dB/km over a 424 nm bandwidth centred at 1504 nm (corresponding to 54.3 THz), and of less than 0.1 dB/km in a 144 nm (17.9 THz) spectral region around 1553 nm.

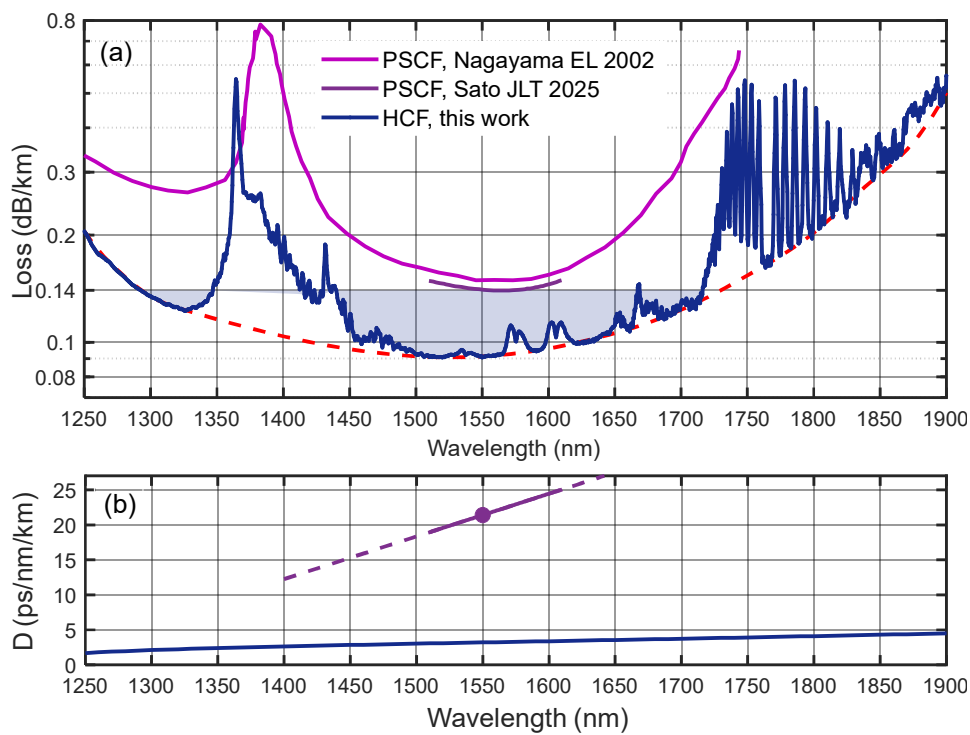

**Fig.3: Loss and chromatic dispersion comparison with state-of-the-art telecoms fibres.** (a) Loss of the DNANF in this work, compared with that of the record-low loss pure silica core fibres (PSCFs) from 2002 (Nagayama et al.[34]) and 2025 (Sato et al.[10]); (b) simulated dispersion of the DNANF and measured dispersion of the PSCF from Sato et al.

Figure 3(b) shows the modelled dispersion of HCF2, with 2.1, 3.2, and 3.7 ps/nm/km at 1310, 1550, and 1700 nm, respectively, and compares it with that of the record low loss PSCF[10]. Not only has HCF2 lower loss and broader bandwidth, but it also presents a seven times reduction in chromatic dispersion (3.2 vs 21.8 ps/nm/km at 1550 nm) and a seventeen times reduction in dispersion slope. For coherent transmissions, this enables simplifications in the transceiver's DSP complexity and energy consumption; for intensity modulation- direct detection used in shorter reach transmission it enables longer transmission lengths without dispersion compensation[35].

## 3. Further improvements and discussion

After discussing the performances of HCF2, we will now extrapolate the optical properties that these types of HCFs could ultimately offer with further engineering work. Figure 4(a) shows once again the loss of state-of-the-art PSCF telecoms fibres. The minimum loss of ~0.14 dB/km is achieved in a spectrally narrow region of around 20 nm around 1550 nm. Approximately 10 THz can be guided in the C and L telecommunication bands with a loss below 0.145 dB/km, and 26 THz can be transmitted with a loss below 0.2 dB/km[9]. The red curves show the fundamental loss (i.e. with gas absorption omitted) of the fabricated HCF2 (dashed) and of an ideal DNANF having the same nominal geometry, cladding and core diameters, with a perfect transverse and longitudinal uniformity. Its loss at 1550 nm is predicted to be 0.07 dB/km and the bandwidth where this remains below 0.2, 0.14, and 0.1 dB/km is 92, 78, and 54 THz, respectively. The figure also shows five other designs, where we have rigidly scaled all the membrane thicknesses by a common factor to shift their antiresonance window at wavelengths around 850, 1060, 1310, 1700, and 2000 nm. In each case, the core size has been optimised so that the total loss contribution from LL, SSL and μBL is minimised (see Methods). As can be seen, by using one of these HCF designs, losses below 0.2 dB/km, compatible with long distance communications, become possible from 700 to over ~2400 nm. This offers the opportunity to optimise the transmission wavelengths of choice based on where optoelectronic components and amplification technologies present the best performance and achieve the lowest cost per bit, as well as the possibility to provide low-loss transmission at wavelengths that have been so far inaccessible. To quantify the opportunity, more than 70 and 105 THz of guidance below 0.1 and 0.14 dB/km, respectively, would be possible for a fibre centred at 1310 nm. Also, 67 and 160 THz below 0.14 and 0.2 dB/km, respectively, would be possible around 1060 nm, and 100 THz below 0.2 dB/km seem achievable around 850 nm. The effective exploitation of bandwidths up to ten times wider than in today's telecom C+L would require suitable wideband amplification technologies. To this extent, the tunability of the low loss transmission wavelength can unlock the possibility to use amplifiers with bandwidth much greater than Erbium in the C-band (4.5 THz), such as Ytterbium at 1060 nm (13.7 THz)[36], Bismuth in the O, E and S bands (21.0 THz)[37], Thulium and Holmium around 2000 nm (31.5 THz)[38] or others[39].

Finally, simulations also indicate a potential route to achieve even lower losses than those reported in Fig. 4(a). As reported in the literature, LL and SSL in HCFs decrease with increasing core sizes, while µBL has the opposite behaviour[29,40]. By making the fibre stiffer (larger glass outer diameter) and with a thicker coating layer to reduce the µBL contribution, the minimum loss of the fibre can reach lower values at larger core sizes. Figure 4(b) illustrates our modelling predictions for a fibre operating at 1550 nm. Starting from the 29.5 µm core diameter of HCF2 (0.07 dB/km modelled if no asymmetries were present), we see that scaling the core to, say, 40 or 50 µm (and by the same ratio also the inter-tube gap sizes and, proportionally, the outer glass diameter and coating thickness), the total loss of the fibre is predicted to decrease to 0.033 and 0.018 dB/km, respectively. Interestingly, even though these fibres would have larger glass tubes than HCF2 with the same membrane thickness, our fluid dynamics model of the fabrication process[41] indicates that their structure are still achievable with standard fabrication methods, with only a modest worsening of their pressure and surface tension-driven dynamics. For example, for a 50 µm core fibre the pressure buffer one would have to avoid detrimental mid-draw contact would reduce by only 25% and the sensitivity of a large tube in the final fibre to externally imposed pressure would increase by 2.7 times. Both seem controllable in practice with extra attention and engineering. Clearly, such fibres would be less suitable for high fibre-count cables; they would be stiffer and less bendable, both from a mechanical and an optical point of view. For example, their critical bend radius would increase from 4.2 to 8.4 and 16 cm, respectively, in the examples above, see Fig. 4(b). Implementing these changes would necessitate significant engineering efforts and adjustments to the existing cabling and installation procedures. However, if the performance benefits are validated, these challenges appear manageable.

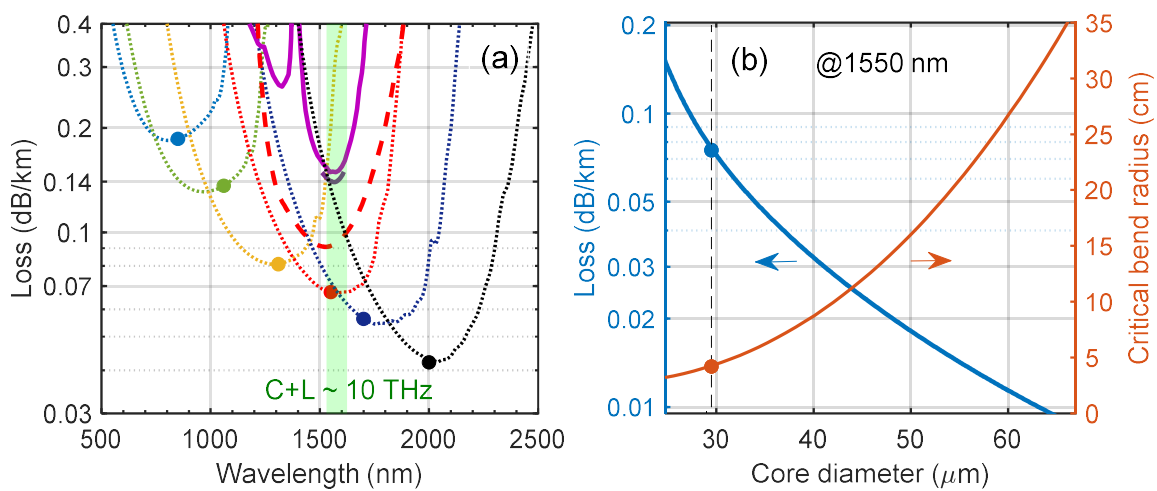

**Fig.4: Simulated potential performance of further optimised fibres.** (a) Loss of state-of-the-art PSCFs (pink and purple, see Fig.2), compared to loss of HCF2 (red dashed, gas absorptions omitted) and ideal DNANF with perfect structure (red dotted). The other dotted curves are modelled losses (LL+SSL+µBL) of DNANFs with core and tube thicknesses optimised for guidance at 850, 1060, 1310, 1700, and 2000 nm; (b) total simulated loss and critical bend radius as a function of core diameter for a 1550 nm fibre similar to HCF2 with suitably larger core diameter and thicker coating.

In conclusion, we have reported what we believe to be one of the most significant improvements in waveguided optical technology for the past 40 years. The pinnacle of six years of improvement and optimisation, the HCF based on a DNANF design reported here offers a completely new paradigm for transmitting data. In addition to a practically negligible optical nonlinearity[31], the capacity to withstand and transmit three to four orders of magnitude higher powers[42], a 30% reduced latency and a six times reduction in chromatic dispersion compared to standard telecom fibres that is typical of these HCFs, our fibre also presents the lowest loss ever measured in an optical waveguide: 0.091 dB/km at 1550 nm and <0.1 dB/km from 1481 to 1625 nm (18 THz). Neglecting absorptions from gases in the core and not of fundamental origin, the fibre guides light with <0.2 dB/km from 1250 to 1730 nm (66 THz), a 250% improvement over current telecoms fibres (25 THz). Our modelling also indicates that DNANF technology offers in principle the opportunity to tune its lowest loss transmission window at arbitrary spectral regions in the near infrared (from 700 to >2000 nm). Here, five to ten times wider bandwidth than the current C+L telecoms bands could be achieved, at spectral locations where existing ultra-broad bandwidth amplifiers currently operate, and at lower losses than fundamentally achievable in silica. Finally, we have shown that losses considerably lower than the value reported in this work and towards the 0.01 dB/km level might be realistically achievable in stiffer fibres with adequately large glass and coated diameters.

In light of the reported results, we are confident that, with advancements in produced volumes, geometrical consistency and reduced presence of absorbing gases in the core, DNANF HCFs will establish themselves as a pivotal waveguiding technology. This innovation has the potential to enable the next technological leap in data communications.

# Methods

## Modelling loss validation

To calibrate and validate our loss models, we acquired SEM images of the end faces of fifteen fabricated DNANFs, each with five sets of nested tubes like the one in Fig. 1(a), but with geometrical parameters (core diameter, tube sizes and membrane thicknesses) varying by up to 5-20%. We then reconstructed the contours of their geometrical structure using the procedure above and performed finite element mode-solving calculations on the resulting permittivity profiles. This solution yields the leakage loss of each mode from the imaginary part of the eigenvalue associated with it. Knowledge of the eigenvalue or mode propagation constants across wavelengths is also used to calculate both the group delay (latency) and chromatic dispersion (such as shown in Fig 3(b)) from the first and second derivatives with respect to optical frequency, respectively. From the mode-field distributions, we computed the normalized field intensity near the glass interfaces $F$, as well as the mode spot area $w_0$ that are used in SSL and µBL models[42]. Here, two unknown parameters are used to empirically characterise the roughness of the glass-air surfaces, ultimately caused by frozen-in surface capillary waves, while two more are used to represent the power spectral density of the external perturbations that lead to microbending. To fit these four parameters, we measured the loss of the 15 different DNANFs. We then extracted their geometry and fitted their calculated total loss, i.e., the sum of LL, SSL, and µBL, to the measured loss curves. We hence obtained a unique set of values for these four non-directly measurable coefficients that ensured the best agreement between measured and total simulated loss for all fibres, simultaneously. One example of the comparison between simulation and measurement is shown in Fig.1(a).

## Fibre fabrication

The fibre fabrication involved a two-stage stack, fuse, and draw technique, where cladding capillaries surrounding a central air core, composed of five outer silica tubes with two sets of five nested middle and inner tubes, were stacked and fused within a jacket tube and then drawn into intermediate size canes. These canes were subsequently reduced to fibre dimensions using a cane-in-tube drawing process. During the fibre drawing, the core region and the three regions between and inside cladding tubes were pressurized independently. The thicknesses of the three sets of five cylindrical silica membranes surrounding the core were engineered to ~500 nm to centre the first antiresonant window to around 1550 nm.

## Distributed loss measurements via Optical Time Domain Reflectometry (OTDR)

The fibre length and the distributed loss of the fibre were measured using a bi-directional OTDR method. One advantage of performing bi-directional measurements is that through simple theory[43] one can separate the longitudinal loss dependence from any local transient change in scattering caused by gas pressure gradients along the fibre[44]. These transients are typically present near both ends of an HCF, due to the ingress of air at atmospheric pressure. In our measurement, the OTDR (Viavi E41DWDMC) was amplified to compensate for the DNANF's 30 dB lower backscattering than in glass-core fibres[45], and coupled into the fibre under test (FUT) through a spliced mode-field adaptor. Thanks to the low optical nonlinearity of DNANF, we were able to boost the pulse energy launched into the fibre using an EDFA, thereby increasing the dynamic range of the measurement. The measurements were performed with a time resolution of 10 to 30 ns, corresponding to a spatial resolution of 1.5 to 4.5 m. Measurements at 1310 and at 1550 nm reported in Fig. 2(b) clearly show uniform loss across the full fibre length.

## Broadband optical attenuation measurement via cutback

The optical attenuation of the fibres was also measured using a cutback technique. The full length of the fibre, measuring 15 km, was spooled on a bobbin of 1 m circumference with a short length of 20 m deployed loosely on the optical bench. Two different sets of equipment (A and B) were employed. Set A involved spectral measurements using an intensity-stabilized tungsten white light source (Bentham WLS100) as a broadband source, and an optical spectrum analyser (OSA, Yokogawa AQ-6315A, wavelength range 400-1750 nm, but rather noisy after 1650 nm). Set B consisted of a white light source, a monochromator and photodetector, integrated within a commercial loss-measurement device. With Set A, the DNANF input end was spliced to a mode-field-adapted solid core patchcord and kept unaltered. Transmission traces from three different cleaves of the DNANF output end were then recorded in the OSA for the full fibre length, then the fibre was cut to 20 m and three more traces corresponding to three different cleaves were acquired. The average transmission for each length was used to calculate the spectral loss trace. We repeated this cutback procedure three times. With set B, we also applied a cutback procedure to acquire one additional loss trace. Here, the launch into the DNANF under test occurred through a free-space mode-field-adapted beam. As this method is virtually independent on the quality of the cleaves, there was no need to acquire multiple cleaves. The four cutbacks returned similar traces, despite the use of different launch and measurement set-ups.

## Mean loss, confidence level and loss measurement at long wavelengths

From the four independent cutback measurements, we calculated the spectral dependence of the mean loss (dark blue curve in Fig. 2(c)) and its standard deviation $\sigma$. The confidence level around the mean value, shown by the green shaded area around the blue line, was calculated as $CE = z^* \cdot \sigma/\sqrt{n}$, with $z^* = 1.96$ for a confidence level of 95% and $n = 4$. We also performed a fifth 20 m cutback using a white light source and a long wavelength OSA (Yokogawa AQ-6375E, wavelength range 1200-2400 nm) to acquire the full spectral loss of the fibre, even beyond the wavelength of 1650 nm where the previous measurement becomes too noisy. This additional measurement is shown by the light blue trace in Fig. 2(c). The whole curve has been translated vertically so that it matched the loss of the more accurate dark blue curve at 1650 nm.

## Intermodal interference measurement

Intermodal interference (IMI) is an impairment in the transmitted signal caused by beating between the signal in the fundamental mode and weak replicas caused by multiple scattering events into the slower higher-order modes and back into the fundamental mode. The incoherent interference between the signal and its weak delayed replica leads to noise, which we measured here using a swept wavelength scanning method. We used a tunable laser source with wavelength resolution of 0.2 picometers and a fast power meter. IMI is calculated using the formula [46]:

$$\alpha^2 = \frac{1}{2}\left(\frac{\sigma}{P_{av}}\right)^2,$$

where $P_{av}$ is the average power of the measurement across a 2 nm window centred at 1555 nm, and $\sigma$ is the standard deviation around it.

## Simulation of wavelength scaling

To investigate the loss and bandwidth that could be potentially achievable in DNANFs similar to HCF2 at other wavelengths, we exploit the principle that in antiresonant HCFs the membrane thickness can be modified for operating at a different wavelength $\lambda$ by ensuring that the normalized frequency $f = \frac{2t\sqrt{n^2-1}}{\lambda}$ remains constant. Here, we calculated the thicknesses required for operation at 850, 1060, 1310, 1700 and 2000 nm. In this study, we used an idealised version of HCF2 (with no asymmetries) and scaled all its membrane thicknesses by the same factor. Next, for each fibre we scaled the core size. We assumed that: 1) our calibrated loss models retain their accuracy across the 700-2400 nm wavelength range, and 2) the fibres maintained the same outer diameter as HCF2. We then used the values for LL, SSL and µBL obtained for 1550 nm and their scaling rules with core size[29] to identify the core diameter that minimised the overall loss at each central wavelength of interest. We then repeated the mode-solving and loss calculations for each optimized fibre. It is important to note that lower losses than those shown in Fig. 4(a) may be potentially achievable, since the simulated structures reported here are based on the structure of HCF2, which reflect our current fabrication processes, rather than the absolute best DNANFs possible.

## Performance extrapolation in fibres with larger core sizes

Next, we investigated whether a further reduction in attenuation may be possible by enlarging the fibre core size. For this study, we only considered the 1550 nm case and started by the ideal HCF2, where SSL is the dominant loss mechanism and LL and µBL only play a small contribution. LL and SSL are known to scale inversely proportionally with the fibre's core diameter. In contrast, µBL increases steeply with the mode spot area, which is directly proportional to the core diameter[42]. It is however well known that µBL can be reduced by improving how the fibre packaging absorbs the external perturbations. For example, this can be achieved by increasing the diameter of the glass to make the fibre stiffer and more resistant to external lateral loads, or by use of coating materials with different mechanical properties, e.g. lower Young's modulus. Here we considered fibres with the same membrane thickness as HCF2 but with increasingly larger core size (and proportionally larger tube diameters and wider inter-tube gaps). As the core enlarged, we increased the total fibre diameter such that the µBL contribution would remain only a small fraction of the total loss. Since LL decreases more rapidly than SSL with increasing core diameter, this study essentially obtained the SSL-limited loss of the fibre. For all fibres, we assumed a coating with the same mechanical properties as HCF2 but with ah thickness proportional to the core size. The result of mode-solving simulations and loss calculations at 1550 nm for fibres with core diameter from 29 to 65 µm is shown in Fig.4(b). One downside of the overall reduction in loss is that as the core diameters increase the fibres become less bendable. To quantify this, we also calculated the bend loss for each fibre as a function of bend radius. To simulate bens loss we used the well-known transformation optics approach, where the distribution of the refractive index *n* in the cross-section is mapped as $n(x,y) \rightarrow n(x,y)e^{\frac{x}{R_b}}$, and $R_b$ is the bend radius. For each fibre, we then determined and plotted the critical bend radius, defined as the radius causing the loss to double from that of the straight fibre.

## Author contributions

## Acknowledgments

The work was partly funded by the ESPRC Prosperity Partnership Programme FASTNET, EP/X025276/1. The authors would like to gratefully acknowledge Shahab Bakhtiari Gorajoobi, Radan Slavík, Xuhao Wei, Marcelo Alonso, Krystian Wisniowski and Leonard Budd for substantial contributions to improving techniques and instruments tailored to the characterisation of HCFs; Alex Boyland, Lucy Hooper, Grzegorz Babiarz, Thejus Varghese, Dmytro Suslov and Mariana Fatobene Ando for contributions to developing and improving manufacturing processes related to HCF production; Jamie Gaudette, David Richardson, Andrew Harker, Andy Appleyard, Stephen Doran, Ghafour Amouzad Mahdiraji and Jaroslaw Rzegocki for constructive discussions related to HCF technology that have contributed indirectly to the result reported in this work.

## Competing interests

The authors declare no competing interests.

## Data availability

The data that support the findings of this study are available from the corresponding author upon reasonable request.